\documentclass[conference,10pt]{IEEEtran}
\usepackage{graphicx}
\usepackage{amssymb}
\usepackage{amsmath}
\usepackage{caption2}

\ifCLASSINFOpdf
\else
\fi

\newtheorem{theorem}{Theorem}

\begin{document}
%
\title{Degrees-of-Freedom Region of Time Correlated MISO Broadcast Channel with Perfect Delayed CSIT and Asymmetric Partial Current CSIT}

\author{\IEEEauthorblockN{Chenxi Hao and Bruno Clerckx}
\IEEEauthorblockA{Communication and Signal Processing Group, Department of Electrical and Electronic Engineering\\
Imperial College London, United Kingdom\\
Email: \{chenxi.hao10,b.clerckx\}@imperial.ac.uk}}

\maketitle

\begin{abstract}
The impact of imperfect CSIT on the degrees of freedom (\emph{DoF})
of a time correlated MISO Broadcast Channel has drawn a lot of
attention recently. Maddah-Ali and Tse have shown that the
completely stale CSIT still benefit the \emph{DoF}. In very recent
works, Yang et al. have extended the results by integrating the
partial current CSIT for a two-user MISO broadcast channel. However,
those researches so far focused on a symmetric case. In this
contribution, we investigate a more general case where the
transmitter has knowledge of current CSI of both users with unequal
qualities. The essential ingredient in our work lies in the way to
multicast the overheard interference to boost the \emph{DoF}. The
optimal \emph{DoF} region is simply proved and its achievability is
shown using a novel transmission scheme assuming an infinite number
of channel uses.
\end{abstract}


\IEEEpeerreviewmaketitle

\section{Introduction}

\vspace{-2pt}Maddah-Ali and Tse have recently investigated the \emph{DoF} region
of Broadcast Channels with perfect delayed (completely stale) CSIT
\cite{Tse10}. They have shown that the per-user optimal \emph{DoF}
is $\frac{2}{3}$ in a two-user setup and proposed a simple
transmission scheme (denoted as MAT scheme in the sequel) to achieve
that \emph{DoF} over 3 time slots. In the first and second slot,
each user in turn receives the desired signal and overhears the
unwanted signal, while in the third slot, the sum of the overheard
interference is transmitted to enable the decoding of the desired
signal at each receiver.
\par Those results have recently been extended to a more general setup
with perfect delayed CSIT and partial current CSIT \cite{Ges12}
\cite{Gou12}. An optimal combining of perfect delayed CSIT and
partial current CSIT is obtained to bridge the \emph{DoF} of
\cite{Tse10} with outdated CSIT and the \emph{DoF} achievable with
zero-forcing beamforming with perfect current CSIT. However results
in \cite{Ges12} and \cite{Gou12} are limited to a symmetric case
where the transmitter has access to both users\rq{} CSI with the
same accuracy.

In \cite{VV11b}, an outer-bound on the \emph{DoF} region was given
under the settings that the transmitter and receivers have multiple
antennas and perfect delayed CSIT. The authors of \cite{VV11a} and
\cite{HJSV09} studied the same settings but without CSIT. Moreover,
In \cite{Maleki}\cite{TMTPS}, the optimal sum \emph{DoF} of
$\frac{3}{2}$ was derived when the CSIT of one user is perfect but
it is out-dated for the other user.

In this paper, we further extend the analysis by considering a more
general asymmetric scenario where the qualities of current CSI of
users\rq{} channels available at the transmitter are unequal. Our
contributions are summarized as follows:
\begin{enumerate}
\item
the transmission schemes derived for the symmetric case are shown to
incur a \emph{DoF} loss in an asymmetric scenario,
\item
the \emph{DoF} region obtained in \cite{Ges12} is extended to the
asymmetric scenario and is expressed as a function of two parameters
representing the accuracy of CSIT of each user's channel,
\item
a novel transmission scheme (that subsumes the scheme in \cite{Ges12} in
the symmetric case) is derived and shown to achieve the \emph{DoF}
region in the limit of an infinite number of channel uses.
\end{enumerate}

At the time of submission, we have been informed of another
independent work, also in submission, that addresses the same problem \cite{Chen12}.
Interestingly, both works derive the same achievable \emph{DoF}
region using different approaches and its achievability is
demonstrated using two different transmission strategies.

The rest of this paper is organized as follows. The system model is
introduced in Section \ref{system_model} and the \emph{DoF} region
is derived in Section \ref{DoF}. The limitations of the transmission
schemes designed for symmetric partial current CSIT are discussed in
Section \ref{achievability} and a novel transmission scheme
achieving the optimal \emph{DoF} region in the setting of asymmetric
partial current CSIT is introduced. Section \ref{conclusions}
concludes the paper.

The following notations are used throughout the paper. Upper letters
in bold font represent matrices whereas bold lower letters stand for
vectors. However, symbol not in bold font represents a scalar.
$\left({\cdot}\right)^T$ and $\left({\cdot}\right)^H$ represent the
transpose and conjugate transpose of a matrix or vector
respectively. $\mathbf{h}^\bot$ denotes the orthogonal space of
channel vector $\mathbf{h}$. $\mathcal{E}\left[{\cdot}\right]$
refers to the expectation of a random variable, vector or matrix.
$\parallel{\cdot}\parallel$ is the norm of a vector.
$f\left({P}\right){\sim}{P^{B}}$ corresponds to
$\lim_{P{\to}{\infty}}\frac{{\log}f\left(P\right)}{{\log}P}{=}B$,
where $P$ refers to SNR throughout the paper and logarithms are in
base $2$.

\section{System Model}\label{system_model}
We consider a two-user Broadcast Channel with two transmit antennas
and one antenna per user. $\mathbf{h}_t$ and $\mathbf{g}_t$ are the
channel states at time slot $t$ of user 1 and user 2 respectively.
Denoting the transmit signal vector as $\mathbf{s}_t$, subject to a
transmit power constraint
$\mathcal{E}\big[\left\|\mathbf{s}_t\right\|^2\big]{\leq}P$, the
observations at receiver 1 and 2, $y_t$ and $z_t$ respectively, can
be written at time slot $t$ as
\begin{align}
y_t & =\mathbf{h}_t^H \mathbf{s}_t + \epsilon_{t,1}\label{eq:model1}\\
z_t & =\mathbf{g}_t^H\mathbf{s}_t+\epsilon_{t,2},\label{eq:model2}
\end{align}where $\epsilon_{t,1}$ and $\epsilon_{t,2}$ are unit power AWGN
noise. Signal vector $\mathbf{s}_t$ is expressed as a function of
the symbol vectors for user 1 and user 2, denoted as $\mathbf{u}_t$
and $\mathbf{v}_t$ respectively. $\mathbf{u}_t$ is a two-element
symbol vector containing $u_{t,1}$ and $u_{t,2}$. $\mathbf{v}_t$ is
defined similarly and is composed of $v_{t,1}$ and $v_{t,2}$. The
power allocated to symbol $u_{t,1}$ is stated as $P_{u_{t,1}}$ while
the rate achieved by $u_{t,1}$ is $R_{u_{t,1}}$. As
$\mathbf{u}_t{=}\left[u_{t,1},u_{t,2}\right]^{T}$, the power and
rate of $\mathbf{u}_t$ are expressed as
$P_{\mathbf{u}_t}{=}P_{u_{t,1}}{+}P_{u_{t,2}}$ and
$R_{\mathbf{u}_t}{=}R_{u_{t,1}}{+}R_{u_{t,2}}$. These notations are
applicable to $\mathbf{v}_t$, $v_{t,1}$ and $v_{t,2}$. For the sake
of convenience, in a few instances, we denote the rate of each
symbol or symbol vector by the pre-log factor (ignoring ${\log}P$).

The channel state information $\mathbf{h}_t$ and $\mathbf{g}_t$ are
supposed to be mutually independent and identically distributed with
zero mean and unit covariance matrix
($\mathcal{E}\left[|\mathbf{h}_t^H\mathbf{g}_t|^2\right]{=}0$ and
$\mathcal{E}\left[\mathbf{h}_t\mathbf{h}_t^H\right]{=}\mathbf{I}_2$).
At any given time slot $t$, the transmitter and each user have
perfect global knowledge of the channel states from time slot $0$ to
$t{-}1$, i.e.\ $\mathbf{h}_{0{,}1{,}{\cdots}{,}t{-}1}$ and
$\mathbf{g}_{0{,}1{,}{\cdots}{,}t{-}1}$. Moreover, the transmitter
can predict the current channel state of each user. The estimated
channels are denoted as $\hat{\mathbf{h}}_t$ and
$\hat{\mathbf{g}}_t$, with the corresponding error vectors written
as $\tilde{\mathbf{h}}_t{=}\mathbf{h}_t{-}\hat{\mathbf{h}}_t$ and
$\tilde{\mathbf{g}}_t{=}\mathbf{g}_t{-}\hat{\mathbf{g}}_t$.
$\tilde{\mathbf{h}}_t$ and $\tilde{\mathbf{v}}_t$ have the variances
denoted as
$\mathcal{E}\big[\parallel\tilde{\mathbf{h}}_t\parallel^2\big]=\sigma_1^2$
and
$\mathcal{E}\left[\parallel\tilde{\mathbf{g}}_t\parallel^2\right]=\sigma_2^2$,
respectively. In the general asymmetric scenario under interest, the
transmitter has unequal accuracy of the current CSI for each user,
with the accuracy of the each current CSIT obtained according to
\begin{equation}
\alpha_k \triangleq \lim_{P\to\infty}-\frac{\log\sigma_k^2}{\log P},k=1,2.\label{eq:alpha}
\end{equation}In other words,
$\mathcal{E}\big[\parallel\tilde{\mathbf{h}}_t\parallel^2\big]{\sim}P^{{-}\alpha_1}$
and
$\mathcal{E}\left[\parallel\tilde{\mathbf{g}}_t\parallel^2\right]{\sim}P^{-\alpha_2}$.
$\alpha_1$ and $\alpha_2$ are supposed to vary in the range of
$\left[0,1\right]$, where $\alpha_k{=}0$ represents no CSIT whereas
$\alpha_k{=}1$ stands for perfect CSIT. Throughout the paper, we
assume $\alpha_1\leq\alpha_2$ without loss of generality. Moreover,
it is important to note the relationship
$\mathcal{E}\left[|\mathbf{h}^H\hat{\mathbf{h}}^\bot|^2\right]{\sim}P^{-\alpha_1}$
and
$\mathcal{E}\left[|\mathbf{g}^H\hat{\mathbf{g}}^\bot|^2\right]{\sim}P^{-\alpha_2}$.

\section{\emph{DoF} Region with Asymmetric Partial Current CSIT}\label{DoF}

\begin{figure}[!t]
\renewcommand{\captionfont}{\small}
\setlength{\abovecaptionskip}{0pt}
\setlength{\belowcaptionskip}{0pt} \captionstyle{center} \centering
\includegraphics[height=7cm,width=9cm]{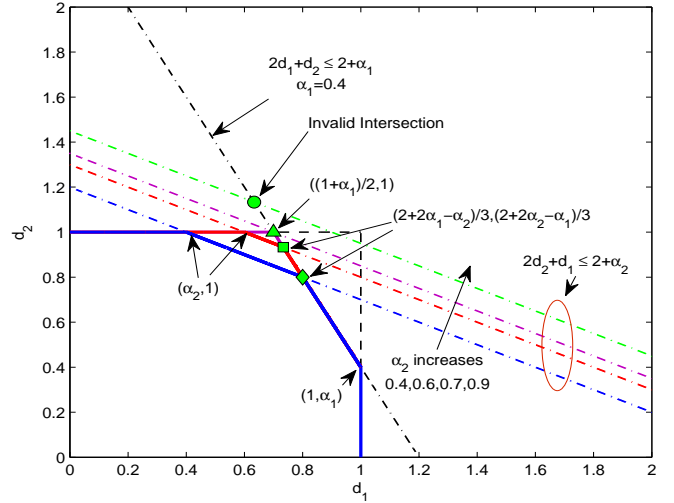}
\caption{DoF region with asymmetric partial current CSIT}\label{fig:dof_reg}
\end{figure}

\begin{theorem} \label{DoF_theorem}
In the two-user MISO Broadcast Channel with perfect delayed CSIT and
asymmetric partial current CSIT, the optimal \emph{DoF} region is
characterized by \setlength{\arraycolsep}{0.2em}
\begin{eqnarray}
\left\{d_1\leq1;d_2\leq1;d_1+2d_2\leq2+\alpha_2;2d_1+d_2\leq2+\alpha_1\right\}.\label{eq:dof_opt}
\end{eqnarray}
\setlength{\arraycolsep}{5pt}
\end{theorem}
A sketch of the proof will be given in the Appendix. The \emph{DoF}
region is illustrated in Figure \ref{fig:dof_reg}. As shown, the
outer-bound is a polygon composed of the points
$\left(1,\alpha_1\right)$, $\left(\alpha_2,1\right)$ and
$\left(\frac{2{+}2\alpha_1{-}\alpha_2}{3},\frac{2{+}2\alpha_2{-}\alpha_1}{3}\right)$.
When $\alpha_1$ is fixed, the intersection point which maximizes the
sum \emph{DoF} moves as $\alpha_2$ increases. However, the
intersection point goes outside the valid region as shown by the
circle point in Figure \ref{fig:dof_reg} when
$2\alpha_2{-}\alpha_1{>}1$. In this case, the region is only formed
by $\left(1,\alpha_1\right)$,
$\left(\frac{1{+}\alpha_1}{2},1\right)$. Also, the point that
achieves the maximum sum \emph{DoF} returns to the triangle point
(see Figure \ref{fig:dof_reg}) achieved by taking
$2\alpha_2{-}\alpha_1{=}1$. When $\alpha_1{=}\alpha_2$, the
\emph{DoF} region boils down to that of \cite{Ges12} as shown by the
diamond point, suggesting that scheme II in \cite{Ges12} achieves a
subset of the asymmetric region. Moreover, when
$\alpha_1{=}\alpha_2{=}0$, the region further boils down to the
region achieved by MAT scheme in \cite{Tse10}.

\section{Achievability}\label{achievability}

\subsection{Limitation of scheme II in
\cite{Ges12}}\label{limitation}

Following the system model defined in Section \ref{system_model}, we briefly review the transmission scheme II as proposed in
\cite{Ges12} and identify the limitations of that scheme in an
asymmetric configuration. The transmit signal and the received signal at each user at first time slot
write as
\begin{align}
\mathbf{s}_1 & = \left[\hat{\mathbf{g}}_1^\bot,\hat{\mathbf{g}}_1\right]\mathbf{u}_1+
\left[\hat{\mathbf{h}}_1^\bot,\hat{\mathbf{h}}_1\right]\mathbf{v}_1,\label{eq:s1}\\
y_1 &= \mathbf{h}_1^H\hat{\mathbf{g}}_1^{\bot}u_{1,1}+ \mathbf{h}_1^H\hat{\mathbf{g}}_1u_{1,2}+\eta_{1,1}+\epsilon_{1,1,}\label{eq:y1}\\
z_1 &= \eta_{1,2}+\mathbf{g}_1^H\hat{\mathbf{h}}_1^{\bot}v_{1,1}+
\mathbf{g}_1^H\hat{\mathbf{h}}_1v_{1,2}+\epsilon_{1,2},\label{eq:z1}
\end{align}where $P_{u_{1,1}}{=}P_{v_{1,1}}{=}P$ and
$P_{u_{1,2}}{=}P^{1{-}\alpha_2}$, $P_{v_{1,2}}{=}P^{1{-}\alpha_1}$.
$\eta_{1,1}$ (resp. $\eta_{1,2}$) represents the overheard
interference generated at receiver 1 (resp. 2) at slot 1 and is
expressed as
\begin{align}
\eta_{1,1} & =
\mathbf{h}_1^H\hat{\mathbf{h}}_1^{\bot}v_{1,1}+ \mathbf{h}_1^H\hat{\mathbf{h}}_1v_{1,2},\\
\eta_{1,2} & = \mathbf{g}_1^H\hat{\mathbf{g}}_1^{\bot}u_{1,1}+
\mathbf{g}_1^H\hat{\mathbf{g}}_1u_{1,2}.
\end{align}By integrating the current CSIT in the precoding procedure to
$\mathbf{u}_1$, $\eta_{1,1}$ and $\eta_{1,2}$ are received with the
power of $P^{1{-}\alpha_1}$ and $P^{1{-}\alpha_2}$ respectively
since $\hat{\mathbf{h}}_1$ and $\hat{\mathbf{g}}_1$ have the
respective quality of $\alpha_1$ and $\alpha_2$. As both overheard
interferences have reduced power, the resource can be saved when
multicasting them separately and sequentially at slot 2 and 3. The
transmission and receive signals at slot 2 and 3 are expressed as
\begin{align}
\mathbf{s}_2 & =\left[\hat{\eta}_{1,1},0\right]^T+\hat{\mathbf{g}}_2^\bot u_2+\hat{\mathbf{h}}_2^\bot v_2,\label{eq:s2}\\
y_2 & =
h_{2,1}^*\hat{\eta}_{1,1}+\mathbf{h}_2^H\hat{\mathbf{g}}_2^{\bot}u_2+\mathbf{h}_2^H\hat{\mathbf{h}}_2^{\bot}v_2+\epsilon_{2,1},\label{eq:y2}\\
z_2 & =
g_{2,1}^*\hat{\eta}_{1,1}+\mathbf{g}_2^H\hat{\mathbf{g}}_2^{\bot}u_2+\mathbf{g}_2^H\hat{\mathbf{h}}_2^{\bot}v_2+\epsilon_{2,2},\label{eq:z2}
\end{align}
\begin{align}
\mathbf{s}_3 & =
\left[\hat{\eta}_{1,2},0\right]^T+\hat{\mathbf{g}}_3^\bot
u_3+\hat{\mathbf{h}}_3^\bot v_3,\label{eq:s3}\\
y_3 & = h_{3,1}^*\hat{\eta}_{1,2}+\mathbf{h}_3^H\hat{\mathbf{g}}_3^{\bot}u_3+
\mathbf{h}_3^H\hat{\mathbf{h}}_3^{\bot}v_3+\epsilon_{3,1},\label{eq:y3}\\
z_3 & =
g_{3,1}^*\hat{\eta}_{1,2}+\mathbf{g}_3^H\hat{\mathbf{g}}_3^{\bot}u_3+\mathbf{g}_3^H\hat{\mathbf{h}}_3^{\bot}v_3+\epsilon_{3,2},\label{eq:z3}
\end{align}where, $\hat{\eta}_{t,1}$ and $\hat{\eta}_{t,2}$ are quantized
versions of overheard interference with the quantization rate of
$1{-}\alpha_1$ and $1{-}\alpha_2$ respectively. $u_2$ and $v_2$ are
new symbols at slot 2 intended to user 1 and user 2 respectively.
Similarly, $u_3$ and $v_3$ are defined at slot 3. The power
allocated to each symbols in \eqref{eq:s2} and \eqref{eq:s3} are
$P_{\hat{\eta}_{1,1}}{=}P_{\hat{\eta}_{1,2}}{=}P$,
$P_{u_2}{=}P_{v_2}{=}P^{\alpha_1}$ and
$P_{u_3}{=}P_{v_3}{=}P^{\alpha_2}$.

At each receiver, $\hat{\eta}_{1,1}$ and $\hat{\eta}_{1,2}$ are
decoded at the first stage of SIC by treating all the other symbols
as noise. At high SNR, $\hat{\eta}_{1,1}$ and $\hat{\eta}_{1,2}$ can
be decoded at both receiver with the same rate as their respective
quantization rate. Then $\hat{\eta}_{1,1}$ (resp.
$\hat{\eta}_{1,2}$) can be employed to cancel the overheard
interference $\eta_{1,1}$ (resp. $\eta_{1,2}$) and provide another
independent observation to $z_1$ (resp. $y_1$). Consequently,
$\mathbf{u}_1$ and $\mathbf{v}_1$ achieve the rate of $2{-}\alpha_2$
and $2{-}\alpha_1$, respectively. At slot 2, $u_2$ (resp. $v_2$) is
drown by the noise in $z_2$ (resp. $y_2$) so that user 1 (resp. user
2) can decode $u_2$ (resp. $v_2$) without interference, thus,
$R_{u_2}{=}R_{v_2}{=}\alpha_1$. At slot 3, the limitation rises
because $\mathbf{h}_3^H\hat{\mathbf{h}}_3^{\bot}v_3$ is overheard by
user 1 with the power of $P^{\Delta}\!$, which is larger than the
noise level. Then, $u_3$ is decoded by treating
$\mathbf{h}_3^H\hat{\mathbf{h}}_3^{\bot}v_3$ as noise so that the
rate of $u_3$ is restricted to $\alpha_1$ while
$R_{v_3}{=}\alpha_2$. To sum up, the \emph{DoF} of each user's
symbols is
$\left(\frac{2{+}2\alpha_1{-}\alpha_2}{3}{,}\frac{2{+}\alpha_2}{3}\right)$.
However, this \emph{DoF} satisfies \eqref{eq:dof_opt} and lies
within the \emph{DoF} region. It is outperformed by the intersection
point
$\left(\frac{2{+}2\alpha_1{-}\alpha_2}{3}{,}\frac{2{+}2\alpha_2{-}\alpha_1}{3}\right)$
and user 2 incurs a \emph{DoF} loss of
$\frac{\Delta{=}\alpha_2{-}\alpha_1}{3}$. The loss vanishes in the
symmetric case as $\alpha_1{=}\alpha_2{=}\alpha$, where the above
scheme achieves the optimal bound \cite{Ges12}.

To boost up the \emph{DoF} achieved at slot 3, the overheard
interference $\mathbf{h}_3^H\hat{\mathbf{h}}_3^{\bot}v_3$ can be
removed via retransmission at slot 4. However, since $v_3$ is
decodable by user 2 at slot 3, this retransmission is wasteful for
user 2. To make the retransmission efficient for both users, we
compose $\mathbf{v}_3$ of $v_{3,1}$ and $v_{3,2}$ as a symbol vector
rather than a single symbol. $v_{3,1}$ is precoded and allocated
with the power same as that of $v_3$, $v_{3,2}$ is transmitted with
power $P^{\Delta}$ so that the power of
$\mathbf{h}_3^H\hat{\mathbf{h}}_3^{\bot}\mathbf{v}_3\!$ is not
enhanced compared to $\mathbf{h}_3^H\hat{\mathbf{h}}_3^{\bot}v_3$.
After receiving and decoding the quantized overheard interference,
user 1 can remove
$\mathbf{h}_3^H\hat{\mathbf{h}}_3^{\bot}\mathbf{v}_3$ while user 2
can decode both $v_{3,1}$ and $v_{3,2}$.

Accordingly, as long as one user overhears interference, the
retransmission of the overheard interference is efficient in
improving the \emph{DoF} if we make the overheard interference
composed of two symbols. Based on this observation, we derive a
novel transmission scheme in the next section.

\subsection{Building Blocks of a New Transmission
Scheme}\label{trans_scheme}

We identify important building blocks of a new transmission scheme
that achieves the asymmetric \emph{DoF} region of Theorem
\ref{DoF_theorem}. Building upon scheme II in \cite{Ges12}, a more
general transmission at a given time slot $t$ writes as
\begin{align}
\mathbf{s}_t &=
\left[\hat{\eta}_{t^\prime},0\right]^T+\left[\hat{\mathbf{g}}_t^\bot,\hat{\mathbf{g}}_t\right]\mathbf{u}_t+
\left[\hat{\mathbf{h}}_t^\bot,\hat{\mathbf{h}}_t\right]\mathbf{v}_t,\label{eq:st}
\end{align}where $\hat{\eta}_{t^\prime}$ can be made up of multiple overheard
interferences generated at any previous slots
${t^\prime}{=}\left\{\tau|\tau{<}t\right\}$. The power and rate
allocated to each symbol are defined for a given quantity $S$ as
shown in Table \ref{tab:pr} that represents a fraction of channel
use and whose physical meaning will appear clearer in the sequel.
\begin{table}[!h]
\renewcommand{\captionfont}{\small}
\captionstyle{center} \centering
\begin{tabular}{ccc}
Symbols & Power & Encoding Rate\\
$\hat{\eta}_{t^\prime}$ & $P-P^S$ & $1-S$\\
$u_1$ & $\frac{P^S}{2}-\frac{P^{S-\alpha_2}}{4}$ & $S$\\
$u_2$ & $\frac{P^{S-\alpha_2}}{4}$ & $S-\alpha_2$\\
$v_1$ & $\frac{P^S}{2}-\frac{P^{S-\alpha_1}}{4}$ & $S$\\
$v_2$ & $\frac{P^{S-\alpha_1}}{4}$ & $S-\alpha_1$\\
\end{tabular}
\caption{Power and rate allocation} \label{tab:pr}
\end{table}Regarding $\mathbf{s}_t$ and the power allocated to
each symbol, the received signal at each user can be written as
\begin{align}
y_t &=
\underbrace{h_{t,1}^*\hat{\eta}_{t^\prime}}_P+\underbrace{\mathbf{h}_t^H\hat{\mathbf{g}}_t^\bot
u_{t,1}}_{P^S}+\underbrace{\mathbf{h}_t^H\hat{\mathbf{g}}_t
u_{t,2}}_{P^{S-\alpha_2}}+\underbrace{\eta_{t,1}}_{P^{S-\alpha_1}}+\epsilon_{t,1},\label{eq:yt}\\
z_t &=
\underbrace{g_{t,1}^*\hat{\eta}_{t^\prime}}_{P}+\underbrace{\eta_{t,2}}_{P^{S-\alpha_2}}+\underbrace{\mathbf{g}_t^H\hat{\mathbf{h}}_t^\bot
v_{t,1}}_{P^S}+\underbrace{\mathbf{g}_t^H\hat{\mathbf{h}}_t
v_{t,2}}_{P^{S-\alpha_1}}+\epsilon_{t,2},\label{eq:zt}
\end{align}where $\eta_{t,1}$ and $\eta_{t,2}$ are overheard interferences
received with the power $P^{S-\alpha_1}$ and $P^{S-\alpha_2}$,
respectively. Both $\eta_{t,1}$ and $\eta_{t,2}$ are composed of two
symbols as
\begin{align}
\eta_{t,1} &= \mathbf{h}_t^H\hat{\mathbf{h}}_t^\bot
v_{t,1}+\mathbf{h}_t^H\hat{\mathbf{h}}_t v_{t,2},\label{eq:eta1}\\
\eta_{t,2} &= \mathbf{g}_t^H\hat{\mathbf{g}}_t^\bot
u_{t,1}+\mathbf{g}_t^H\hat{\mathbf{g}}_t u_{t,2}.\label{eq:eta2}
\end{align}To save the
resource required by retransmission, $\eta_{t,1}$ and $\eta_{t,2}$
are quantized at the end of slot $t$ as
\begin{equation}
\eta_{t,1}=\hat{\eta}_{t,1}+\tilde{\eta}_{t,1},\quad
\eta_{t,2}=\hat{\eta}_{t,2}+\tilde{\eta}_{t,2}, \label{eq:q}
\end{equation}where, $\hat{\eta}_{t,1}$ and $\hat{\eta}_{t,2}$ are the quantized
overheard interference, ${\tilde{\eta}_{t,1}}$ and
${\tilde{\eta}_{t,2}}$ are the quantization errors. According to the
Rate-distortion theory \cite{Inf_Theo}, we quantize $\eta_{t,1}$ and
$\eta_{t,2}$ using $R_{\eta_{t,1}}$ and $R_{\eta_{t,2}}$ bits, which
are equal to
$\left(1{-}\alpha_1\right){\log}P{+}o\left({\log}P\right)$ and
$\left(1{-}\alpha_2\right){\log}P{+}o\left({\log}P\right)$
respectively, allowing the quantization noise to have the same
variances as AWGN. As a result, $\hat{\eta}_{t,1}$ and
$\hat{\eta}_{t,2}$ can be employed to remove the overheard
interference completely while providing an additional observation to
enable the decoding of all symbols.

Moreover, $\hat{\eta}_{t^\prime}$, decoded with rate $1{-}S$ by
treating all the other components as noise at both receivers, can be
considered as occupying $1{-}S$ channel use, leaving $S$ channel use
for the new symbols, $\mathbf{u}_t$ and $\mathbf{v}_t$. The number
of new symbols transmitted highly depends on the value of $S$. If
$S{>}\alpha_2$, both users will observe an overheard interference.
When $\alpha_1{<}S{\leq}\alpha_2$, only $u_{t,1}$ is sent to user 1
and will be drown by the noise in $z_t$, thus, only one overheard
interference ($\eta_{t,1}$) needs to be retransmitted. However, when
$S{<}\alpha_1$, only $u_{t,1}$ and $v_{t,1}$ are transmitted and
they will be drown by the noise in $z_t$ and $y_t$ respectively so
that the transmission can finalize without the requirement of
overheard interference retransmission.

Looking at the aforementioned transmission model, all the new
symbols are not decodable until the transmissions of overheard
interferences are completed. As a consequence, we consider all the
channel uses employed to decode all the new symbols as a ''virtual
channel'', which normally consists of two parts: the first part is
the transmission of new symbols, the second part refers to the
retransmission of overheard interferences, which can boost the
\emph{DoF} from two aspects: 1) help user remove overheard
interference; 2) provide an additional independent observation to
enable decoding.

Next, as the limitation of the scheme in \cite{Ges12} lies at time
slot 3, we derive the achievable scheme maintaining the transmission
in the first two time slots plus the transmission of $\eta_{1,2}$ at
time slot 3. Over those $3{-}\alpha_2$ channel uses, the \emph{DoF}
achieved by each user is
$\left(d_1{,}d_2\right){\!=\!}\left(\frac{d_{\mathbf{u}_1}{+}d_{u_2}}{3{-}\alpha_2}{,}
\frac{d_{\mathbf{v}_1}{+}d_{v_2}}{3{-}\alpha_2}\right){\!\!=\!\!}
\left(\frac{2{+}\alpha_1{-}\alpha_2}{3{-}\alpha_2}{,}\frac{2}{3{-}\alpha_2}\right)$.

Moreover, as illustrated in Figure \ref{fig:dof_reg}, the \emph{DoF}
region varies depending on the location of the intersection point.
Obviously, when $2\alpha_2{-}\alpha_1{\geq}1$, the intersection
point lies on or outside the valid region formed by $d_1{=}1$ and
$d_2{=}1$. The maximum sum \emph{DoF} is obtained by the point
$\left(\!\frac{1{+}\alpha_1}{2},1\!\right)$. When
$2\alpha_2{-}\alpha_1{<}1$, the intersection point is located inside
the valid region and achieves the maximum sum \emph{DoF}. Hence, the
achievable schemes are sketched in these two cases respectively.

\subsection{Case I: $1-\Delta\leq\alpha_2$-Achieving $\left(\frac{1+\alpha_1}{2},1\right)$}\label{case1}

\textbf{Lemma 1}: Assume the perfect delayed CSIT and partial
current CSIT with $\alpha_1$ and $\alpha_2$ for each user. Under the
condition $2\alpha_2{-}\alpha_1{\geq}1$
($1{-}\Delta{\leq}\alpha_2$), the achievable \emph{DoF} is
$\left(d_1{,}d_2\right){=}\left(\frac{1{+}\alpha_1}{2}{,}1\right)$.

\emph{Proof}: In addition to the existing two time slots of scheme
II in \cite{Ges12} shown as \eqref{eq:s1} and \eqref{eq:s2}, at time
slot 3, the transmission is processed according to the scheme
described in section \ref{trans_scheme}, which is expressed in the
following group of equations as
\begin{align}
\mathbf{s}_3 & =
\left[\hat{\eta}_{1,2},0\right]^T+\hat{\mathbf{g}}_3^{\bot}u_3+\left[\hat{\mathbf{h}}_3^{\bot},\hat{\mathbf{h}}_3\right]
\mathbf{v}_3,\label{eq:s3I}\\
\mathbf{s}_{4} & =
\left[\hat{\eta}_{3,1},0\right]^T+\hat{\mathbf{g}}_{4}^{\bot}u_4+\left[\hat{\mathbf{h}}_4^{\bot},\hat{\mathbf{h}}_4\right]
\mathbf{v}_4,\label{eq:s4I}\\
\mathbf{s}_5 & =
\left[\hat{\eta}_{4,1},0\right]^T+\hat{\mathbf{g}}_5^{\bot}u_5+\left[\hat{\mathbf{h}}_5^{\bot},\hat{\mathbf{h}}_5\right]
\mathbf{v}_5, \label{eq:s5I}
\end{align}with the power and rate allocation given in Table
\ref{tab:pr1}, where $\Delta{=}\alpha_2{-}\alpha_1$.
\begin{table}[!h]
\renewcommand{\captionfont}{\small}
\captionstyle{center} \centering
\begin{tabular}{ccc}
Symbols & Power & Encoding Rate\\
$\hat{\eta}_{1,2}$ & $P-P^{\alpha_2}$ & $1-\alpha_2$\\
$u_3$ & $\frac{P^{\alpha_2}}{2}$ & $\alpha_2$\\
$v_{3,1}$ & $\frac{P^{\alpha_2}}{2}-\frac{P^\Delta}{4}$ & $\alpha_2$\\
$v_{3,2}$ & $\frac{P^{\Delta}}{4}$ & $\Delta$\\
$\hat{\eta}_{3,1}$ & $P-P^{1-\Delta}$ & $\Delta$\\
$u_4$ & $\frac{P^{1-\Delta}}{2}$ & $1-\Delta$\\
$v_{4,1}$ & $\frac{P^{1-\Delta}}{2}-\frac{P^{{1-\Delta}-\alpha_1=1-\alpha_2}}{4}$ & $1-\Delta$\\
$v_{4,2}$ & $\frac{P^{1-\alpha_2}}{4}$ & $1-\alpha_2$\\
$\hat{\eta}_{4,1}$ & $P-P^{\alpha_2}$ & $1-\alpha_2$\\
\end{tabular}
\caption{Power and rate allocation for case I.} \label{tab:pr1}
\end{table}The power of $u_5$ and $\mathbf{v}_5$ are not shown because
$P_{u_5}{=}P_{u_3}$, $P_{v_{5,1}}{=}P_{v_{3,1}}$ and
$P_{v_{5,2}}{=}P_{v_{3,2}}$. As in \eqref{eq:s4I} and
\eqref{eq:s5I}, the transmission signal have the same form, the
received signal at each user can be written in general as
\begin{align}
y_t &=
\underbrace{h_{t,1}^*\hat{\eta}_{t-1,1}}_P+\underbrace{\mathbf{h}_t^H\hat{\mathbf{g}}_t^\bot
u_t}_{P_{u_t}}+\underbrace{\eta_{t,1}}_{P_{\mathbf{v}_t}P^{-\alpha_1}}+\epsilon_{t,1},\label{eq:ytI}\\
z_t &=
\underbrace{g_{t,1}^*\hat{\eta}_{t-1,1}}_P+\underbrace{\mathbf{g}_t^H\hat{\mathbf{h}}_t^\bot
v_{t,1}}_{P_{v_{t,1}}}+\underbrace{\mathbf{g}_t^H\hat{\mathbf{h}}_t
v_{t,2}}_{P_{v_{t,2}}}+\epsilon_{t,2},\label{eq:ztI}
\end{align}where $\eta_{t,1}$ is given as \eqref{eq:eta1} and quantized as
$\hat{\eta}_{t,1}$ via \eqref{eq:q}. Regarding the transmission flow
and power allocation, two virtual channels can be established from
slot 3 to 5.
\subsubsection{Virtual Channel 1}This virtual channel contains the
transmission of $u_3$, $\mathbf{v}_3$ plus $\hat{\eta}_{3,1}$. At
slot 3, $\alpha_2$ channel use remains after transmitting
$\hat{\eta}_{1,2}$, so that one symbol, $u_3$, is intended to user 1
and two symbols, $v_{3,1}$ and $v_{3,2}$ to user 2. The interference
overheard by user 1, $\eta_{3,1}$, has power
$P_{\mathbf{v}_3}P^{{-}\alpha_1}{=}P^{\Delta}$, while user 2
overhears nothing since $u_3\!$ is drown by the noise. At the end of
slot 3, $\eta_{3,1}$ is quantized as $\hat{\eta}_{3,1}$ with rate
$R_{\eta_{3,1}}$ and sent using $\Delta$ channel use at slot 4.
\subsubsection{Virtual Channel 2}This virtual channels consists of
the transmission of $u_4$ and $\mathbf{v}_4$ together with
$\hat{\eta}_{4,1}$. At slot 4, $1{-}\Delta$ channel remains for
$u_4$ and $\mathbf{v}_4$ after transmitting $\hat{\eta}_{3,1}$.
Under the condition that $1{-}\Delta{\leq}\alpha_2$, the transmitter
will send only one symbol, $u_4$ to user 1 but two symbols,
$v_{4,1}$ and $v_{4,2}$ to user 2. Consequently, $\eta_{4,1}$ is the
only overheard interference at slot 4 with power $P^{1{-}\alpha_2}$.
In each virtual channel, three new symbols and one overheard
interference are transmitted. The only difference between these two
virtual channels lies in the power allocation. The decodability is
sketched next.

Firstly, as aforementioned, the overheard interference must be
decoded first to enable the decoding procedure for the new symbols.
From $y_{t{+}1}$ and $z_{t{+}1}$, $\hat{\eta}_{t,1}$ is decoded by
treating all the other symbols as noise. To be specific, at
observation $y_{t{+}1}$, the rate of $\hat{\eta}_{t,1}$ is
$I\left(\hat{\eta}_{t,1};y_{t{+}1}|h_{t{+}1,1}^*\right){=}{\log}\frac{P}{P_{u_{t{+}1}}}$,
while it is
$I\left(\hat{\eta}_{t,1};z_{t{+}1}|g_{t{+}1,1}^*\right){=}{\log}\frac{P}{P_{v_{t{+}1,1}}}$
from $z_{t{+}1}$. As shown in Table \ref{tab:pr1}, we let
$P_{u_{t{+}1}}{=}P_{v_{t{+}1,1}}$, $\hat{\eta}_{t,1}$ is decoded
with the same rate at both receivers. Moreover, as $\eta_{t,1}$ is
seen by user 1 with power of $P_{\mathbf{v}_t}P^{{-}\alpha_1}$ in
\eqref{eq:ytI}, $\hat{\eta}_{t,1}$ can completely recover
$\eta_{t,1}$ provided that
$\frac{P}{P_{u_{t{+}1}}}{=}P_{\mathbf{v}_t}P^{{-}\alpha_1}$.

Secondly, $\hat{\eta}_{t,1}$ is employed to remove the overheard
interference in \eqref{eq:ytI} and provide an additional independent
observation for user 2. Denoting
$y_t^\prime{=}y_t{-}h_{t,1}^*\hat{\eta}_{t{-}1,1}$ and
$z_t^\prime{=}z_t{-}g_{t,1}^*\hat{\eta}_{t{-}1,1}$ as the signal
after decoding and subtracting $\hat{\eta}_{t{-}1,1}$ from $y_t$ and
$z_t$ respectively, we write the decoding of $\mathbf{v}_t$ and
$u_t$ as
\begin{align}
y_t^\prime-\hat{\eta}_{t,1} &= \mathbf{h}_t^H\hat{\mathbf{g}}_t^\bot u_t+\tilde{\eta}_{t,1}+\epsilon_{t,1},\label{eq:dec_ua}\\
\left[\begin{array}{c}z_t^\prime\\ \hat{\eta}_{t,1}\end{array}\right] &= \left[\begin{array}{c}\mathbf{g}_t^H\\
\mathbf{h}_t^H\end{array}\right]\left[\hat{\mathbf{h}}_t^\bot,\hat{\mathbf{h}}_t\right]\mathbf{v}_t+
\left[\!\!\begin{array}{c}\epsilon_{t,2}\\-\tilde{\eta}_{t,1}\end{array}\!\!\right],\label{eq:dec_va}
\end{align}
where $\tilde{\eta}_{t,1}$, $\epsilon_{t,1}$ and $\epsilon_{t,2}$
have unit power so that $u_t$ is decodable with the rate of
${\log}P_{u_t}$ and $\mathbf{v}_t$ achieves the rate of
${\log}\left(P_{v_{t,1}}P_{v_{t,2}}\right)$.

Replacing the power stated in Table \ref{tab:pr1}, the first virtual
channel lasts for $\alpha_2{+}\Delta$ channel uses, over which the
rates achieved are $R_{u_3}{=}\alpha_2$ and
$R_{\mathbf{v}_3}{=}\alpha_2{+}\Delta$. Similarly, they are
$R_{u_4}{=}1{-}\Delta$ and $R_{\mathbf{v}_4}{=}2{-}\alpha_2-\Delta$
during the $2{-}\alpha_2{-}\Delta$ channel uses in virtual channel
2. In all, when we consider these two virtual channels which last
for 2 channel uses, the \emph{DoF} is
\begin{eqnarray}
\left(d_1,d_2\right)=\left(\frac{R_{u_3}+R_{u_4}}{2},\frac{R_{\mathbf{v}_3}+R_{\mathbf{v}_4}}{2}\right)=\left(\frac{1+\alpha_1}{2},1\right).
\end{eqnarray}
Moreover, when the transmission continues, the amount of remaining
channel uses at slot 5 is $\alpha_2$, which is identical to slot 3.
Repeating the same transmission results in $\eta_{5,1}$ overheard by
user 1 with the rate $R_{\eta_{5,1}}{=}R_{\eta_{3,1}}$, leading to
$P_{u_6}{=}P_{u_4}$ and $P_{\mathbf{v}_6}{=}P_{\mathbf{v}_4}$ at
slot 6. In this way, the transmission can keep cycling with the same
strategy as at slot 3 and 4. Combining with the transmissions prior
$u_3$ and $\mathbf{v}_3$ and assume the strategy is repeated for $N$
times, the asymmetric \emph{DoF} is given by
\setlength{\arraycolsep}{0.2em}
\begin{eqnarray}
\left(d_1,d_2\right)& =& \lim_{N\to\infty}\left(
\frac{2+\alpha_1-\alpha_2+N\times\left(1+\alpha_1\right)}{3-\alpha_2+N\times2},\cdots\right.\nonumber\\
& & \left.\frac{2+N}{3-\alpha_2+N\times2}\right)= \left(\frac{1+\alpha_1}{2},1\right).\label{eq:ds1}
\end{eqnarray}
\setlength{\arraycolsep}{5pt} $\hfill\Box$

\subsection{Case II: $1-\Delta>\alpha_2$-Achieving
$\left(\frac{2+2\alpha_1-\alpha_2}{3},\frac{2+2\alpha_2-\alpha_1}{3}\right)$}\label{case2}

\textbf{Lemma 2}: Assume the transmitter has perfect delayed CSIT
and partial current CSIT with $\alpha_1$ and $\alpha_2$ for each
user. Under the condition $2\alpha_2{-}\alpha_1{<}1$
($1{-}\Delta{>}\alpha_2$), the achievable \emph{DoF} is
$\left(d_1{,}d_2\right){=}\left(\frac{2{+}2\alpha_1{-}\alpha_2}{3}{,}\frac{2{+}2\alpha_2{-}\alpha_1}{3}\right)$.

\emph{Proof}: To close the gap mentioned in section
\ref{limitation}, we derive the achievable scheme from slot 3 (The
first two slots are the same as in Case I). The transmission flow is
expressed in the following group of equations as
\begin{align}
\mathbf{s}_3 & =
\left[\hat{\eta}_{1,2},0\right]^T+\hat{\mathbf{g}}_3^{\bot}u_3+\left[\hat{\mathbf{h}}_3^{\bot},\hat{\mathbf{h}}_3\right]
\mathbf{v}_3\label{eq:s3_2},\\
\mathbf{s}_4 & =
\left[\hat{\eta}_{3,1},0\right]^T+\left[\hat{\mathbf{g}}_4^{\bot},\hat{\mathbf{g}}_4\right]\mathbf{u}_4+
\left[\hat{\mathbf{h}}_4^{\bot},\hat{\mathbf{h}}_4\right]\mathbf{v}_4,\\
\mathbf{s}_5 & =
\left[\hat{\eta}_{4,1},0\right]^T+\hat{\mathbf{g}}_5^{\bot}u_5+\left[\hat{\mathbf{h}}_5^{\bot},\hat{\mathbf{h}}_5\right]\mathbf{v}_5,\\
\mathbf{s}_6 & =
\left[\hat{\eta}_{4,2},0\right]^T+\left[\hat{\eta}_{5,1},0\right]^T+\hat{\mathbf{g}}_6^{\bot}u_6+
\left[\hat{\mathbf{h}}_6^{\bot},\hat{\mathbf{h}}_6\right]
\mathbf{v}_6,
\end{align}where, $\eta_{t,1}$ and $\eta_{t,2}$, given in
\eqref{eq:eta1} and \eqref{eq:eta2}, are quantized with rate
$R_{\eta_{t,1}}$ and $R_{\eta_{t,2}}$ respectively at the end of
each slot via \eqref{eq:q}. Since the transmission and power
allocation at slot 3 and 5 are identical to that in Table
\ref{tab:pr1}, Table \ref{tab:pr2} only provides the power and the
encoding rate at slot 4 and 6.
\begin{table}[h]
\renewcommand{\captionfont}{\small}
\captionstyle{center} \centering
\begin{tabular}{ccc}
Symbols & Power & Encoding Rate\\
$\hat{\eta}_{3,1}$ & $P-P^{1-\Delta}$ & $\Delta$\\
$u_{4,1}$ & $\frac{P^{1-\Delta}}{2}-\frac{P^{1-\Delta-\alpha_2}}{4}$ & $1-\Delta$\\
$u_{4,2}$ & $\frac{P^{1-\Delta-\alpha_2}}{4}$ & $1-\Delta-\alpha_2$\\
$v_{4,1}$ & $\frac{P^{1-\Delta}}{2}-\frac{P^{1-\alpha_2}}{4}$ & $1-\Delta$\\
$v_{4,2}$ & $\frac{P^{1-\alpha_2}}{4}$ & $1-\alpha_2$\\
$\hat{\eta}_{4,2}$ & $P-P^{\Delta+\alpha_2}$ & $1-\Delta-\alpha_2$\\
$\hat{\eta}_{5,1}$ & $P^{\Delta+\alpha_2}-P^{\alpha_2}$ & $\Delta$\\
$u_6$ & $\frac{P^{\alpha_2}}{2}$ & $\alpha_2$\\
$v_{6,1}$ & $\frac{P^{\alpha_2}}{2}-\frac{P^{\Delta}}{4}$ & $\alpha_2$\\
$v_{6,2}$ & $\frac{P^{\Delta}}{4}$ & $\Delta$
\end{tabular}
\caption{Power and rate allocation for case II (slot 4 and 6).}
\label{tab:pr2}
\end{table}

Regarding the transmission flow and power allocation, three virtual
channels are constructed from slot 3 to 6.
\subsubsection{Virtual Channel 1}
This virtual channel consists of the transmission of $u_3$,
$\mathbf{v}_3$ and $\hat{\eta}_{3,1}$. Power of $u_3$ and
$\mathbf{v}_3$ are stated in Table \ref{tab:pr1}. $\mathbf{v}_3$
results in $\eta_{3,1}$, which is quantized as $\hat{\eta}_{3,1}$.
The sending of $\hat{\eta}_{3,1}$ occupies $\Delta$ channel uses at
slot 4. This virtual channel is identical to the first virtual
channel in case I (see Section \ref{case1}). The decodability can be
derived in the same way as \eqref{eq:ytI} and \eqref{eq:ztI}. The
total amount of channel uses is $\alpha_2{+}\Delta$, over which the
rates achieved by the symbols for each user are $R_{u_3}{=}\alpha_2$
and $R_{\mathbf{v}_3}{=}\alpha_2{+}\Delta$.
\subsubsection{Virtual Channel 2}
This virtual channel is made up of $\mathbf{u}_4$ and $\mathbf{v}_4$
and the retransmission of $\hat{\eta}_{4,1}$ and $\hat{\eta}_{4,2}$.
At slot 4, the remaining amount of channel use for new symbols is
$1{-}\Delta$, which is higher than $\alpha_2$ so that the
transmitter sends two symbols per user, resulting in the overheard
interference $\eta_{4,1}$ and $\eta_{4,2}$ generated at user 1 and
user 2 respectively. The observations at each receiver are written
as
\begin{align}
y_4 &=
\underbrace{h_{4,1}^*\hat{\eta}_{3,1}}_P+\underbrace{\mathbf{h}_4^H\hat{\mathbf{g}}_4^\bot
u_{4,1}}_{P_{u_{4,1}}}+\underbrace{\mathbf{h}_4^H\hat{\mathbf{g}}_4
u_{4,2}}_{P_{u_{4,2}}}+\!\!\!\underbrace{\eta_{4,1}}_{P_{\mathbf{v}_4}P^{-\alpha_1}}\!\!\!+\epsilon_{4,1},\label{eq:ytII}\\
z_4&=
\underbrace{g_{4,1}^*\hat{\eta}_{3,1}}_P+\!\!\!\underbrace{\eta_{4,2}}_{P_{\mathbf{u}_4}P^{-\alpha_2}}\!\!\!+
\underbrace{\mathbf{g}_4^H\hat{\mathbf{h}}_4^\bot
v_{4,1}}_{P_{v_{4,1}}}+\underbrace{\mathbf{g}_4^H\hat{\mathbf{h}}_4
v_{4,2}}_{P_{v_{4,2}}}+\epsilon_{4,2}.\label{eq:ztII}
\end{align}At the end of slot 4, $\hat{\eta}_{4,1}$ and $\hat{\eta}_{4,2}$ are
obtained via quantization with rates
${\log}\left(P_{\mathbf{v}_4}P^{{-}\alpha_1}\right)$ and
${\log}\left(P_{\mathbf{u}_4}P^{{-}\alpha_2}\right)$ respectively,
whose pre-log factors are $1{-}\alpha_2$ and
$1{-}\Delta{-}\alpha_2$. The decodability of the new symbols in this
virtual channel is enabled after decoding $\hat{\eta}_{4,1}$ and
$\hat{\eta}_{4,2}$, which are transmitted using part of the channel
at slot 4 and 5 respectively. The received signals of user 1 at slot
5 and 6 are expressed as
\begin{align}
y_5 &=
\underbrace{h_{5,1}^*\hat{\eta}_{4,1}}_P+\underbrace{\mathbf{h}_5^H\hat{\mathbf{g}}_5^\bot
u_5}_{P_{u_5}}+\underbrace{\eta_{5,1}}_{P_{\mathbf{v}_5}P^{-\alpha_1}}+\epsilon_{5,1},\\
y_6 &=
\underbrace{h_{6,1}^*\hat{\eta}_{4,2}}_P+\underbrace{h_{6,1}^*\hat{\eta}_{5,1}}_{P^{\Delta+\alpha_2}}+
\underbrace{\mathbf{h}_6^H\hat{\mathbf{g}}_6^\bot
u_6}_{P_{u_6}}+\underbrace{\eta_{6,1}}_{P_{\mathbf{v}_6}P^{-\alpha_1}}+\epsilon_{6,1}.\label{eq:y6}
\end{align}By treating all the other symbols as noise, $\hat{\eta}_{4,1}$ is
obtained with the rate of
$I\left(\hat{\eta}_{4,1};y_5|h_{5,1}^*\right){=}{\log}\frac{P}{P_{u_5}}$.
Revisiting Table \ref{tab:pr1}, $P_{u_5}{=}P_{u_3}{=}P^{\alpha_2}$
so that $\hat{\eta}_{4,1}$ is decodable with rate $1{-}\alpha_2$ and
$\eta_{4,1}$ can be successfully recovered. Similarly,
$\hat{\eta}_{4,2}$ is decoded from $y_6$ by treating all the other
symbols as noise, among which, $\hat{\eta}_{5,1}$ is the dominant
component. Consequently, $\hat{\eta}_{4,2}$ is decoded with the rate
of
$I\left(\hat{\eta}_{4,2};y_6|h_{6,1}^*\right){=}{\log}\frac{P}{P_{\hat{\eta}_{5,1}}}$,
which meets its distortion rate $1{-}\Delta{-}\alpha_2$. Similarly,
$\hat{\eta}_{4,1}$ and $\hat{\eta}_{4,2}$ are decoded from $z_5$ and
$z_6$ respectively. Denoting
$y_4^\prime{=}y_4{-}h_{4,1}^*\hat{\eta}_{3,1}$ and
$z_4^\prime{=}z_4{-}g_{4,1}^*\hat{\eta}_{3,1}$, the decoding
formulas for $\mathbf{u}_4$ and $\mathbf{v}_4$ are
\begin{align}
\left[\begin{array}{c}y_4^\prime-\hat{\eta}_{4,1}\\ \hat{\eta}_{4,2}\end{array}\right]&{=}\left[\begin{array}{c}\mathbf{h}_4^H\\
\mathbf{g}_4^H\end{array}\right]\left[\hat{\mathbf{g}}_4^\bot,\hat{\mathbf{g}}_4\right]\mathbf{u}_4\!{+}\!
\left[\!\begin{array}{c}\tilde{\eta}_{4,1}+\epsilon_{4,1}\\-\tilde{\eta}_{4,2}\end{array}\!\!\right],\label{eq:dec_ub}\\
\left[\begin{array}{c}z_4^\prime-\hat{\eta}_{4,2}\\ \hat{\eta}_{4,1}\end{array}\right]&{=}\left[\begin{array}{c}\mathbf{g}_4^H\\
\mathbf{h}_4^H\end{array}\right]\left[\hat{\mathbf{h}}_4^\bot,\hat{\mathbf{h}}_4\right]\mathbf{v}_4\!{+}\!
\left[\!\begin{array}{c}\tilde{\eta}_{4,2}+\epsilon_{4,2}\\-\tilde{\eta}_{4,1}\end{array}\!\!\right],\label{eq:dec_vb}
\end{align}where $\tilde{\eta}_{4,1}$, $\tilde{\eta}_{4,2}$, $\epsilon_{4,1}$
and $\epsilon_{4,2}$ have unit power, $\mathbf{u}_4$ and
$\mathbf{v}_4$ can be decoded with the rate of
${\log}P_{\mathbf{u}_4}$ and ${\log}P_{\mathbf{v}_4}$, respectively,
which equal their encoding rate. As a result, virtual channel 2
lasts for $3{-}2\alpha_2{-}2\Delta$ channel uses while the rate
achieved are $R_{\mathbf{u}_4}{=}2{-}2\Delta{-}\alpha_2$ and
$R_{\mathbf{v}_4}{=}2{-}\Delta{-}\alpha_2$.
\subsubsection{Virtual Channel 3}
This virtual channel contains the transmissions of $u_5$,
$\mathbf{v}_5$ and $\hat{\eta}_{5,1}$. As the transmission and power
allocation of $u_5$ and $\mathbf{v}_5$ are identical to that at slot
3, $\eta_{5,1}$ is quantized with the rate $\Delta$. After
subtracting $\hat{\eta}_{4,2}$ from \eqref{eq:y6},
$\hat{\eta}_{5,1}$ is decoded by treating the rest as noise, among
which, $u_6$ is the dominant. The decoding rate of
$\hat{\eta}_{5,1}$ is
$I\left(\hat{\eta}_{5,1};y_6|h_{6,1}^*,\hat{\eta}_{4,2}\right){=}{\log}\frac{P_{\hat{\eta}_{5,1}}}{P_{u_6}}$,
whose pre-log factor is $\Delta$. Consequently, $\hat{\eta}_{5,1}$
can be employed to remove the interference in $y_5$ and provide an
additional observation to $z_5$. In this way, $u_5$ and
$\mathbf{v}_5$ can be decoded using \eqref{eq:ytI} and
\eqref{eq:ztI}. The rates achieved are $R_{u_5}{=}\alpha_2$ and
$R_{\mathbf{v}_5}{=}\alpha_2{+}\Delta$ over the $\alpha_2{+}\Delta$
channel use.

In all, looking at those three virtual channels which last for 3
channel uses in total, the \emph{DoF} achieved are expressed as
\begin{align}
\left(d_1,d_2\right) & = \frac{\left(R_{u_3}+R_{\mathbf{u}_4}+R_{u_5},R_{\mathbf{v}_3}+
R_{\mathbf{v}_4}+R_{\mathbf{v}_5}\right)}{2\left(\alpha_2+\Delta\right)+3-2\alpha_2-2\Delta}\nonumber\\
& =
\left(\frac{2+2\alpha_1-\alpha_2}{3},\frac{2+2\alpha_2-\alpha_1}{3}\right).
\end{align}Combining with the \emph{DoF} prior
to $u_3$ and $\mathbf{v}_3$ and cycling those three virtual channels
for $N$ times, the achievable \emph{DoF} is
\begin{multline}
\left(d_1,d_2\right)=\lim_{N\to\infty}\left(
\frac{2+\alpha_1-\alpha_2+N\times3d_1}{3-\alpha_2+N\times3},\cdots\right.\nonumber\\
\left.\frac{2+N\times3d_2}{3-\alpha_2+N\times3}\right)= \left(\frac{2+2\alpha_1-\alpha_2}{3},\frac{2+2\alpha_2-\alpha_1}{3}\right).\label{eq:ds2}
\end{multline}$\hfill\Box$

\subsection{Case II: $1{-}\Delta{>}\alpha_2$-Achieving $\left(\alpha_2,1\right)$}

Point $\left(\alpha_2,1\right)$ is achieved under the condition that
$1{-}\Delta{>}\alpha_2$ as in case I but by reusing the flow
described in Section \ref{case1}. The first virtual channel is
maintained, where $u_3$ and $\mathbf{v}_3$ achieve the rate of
$\alpha_2$ and $\alpha_2{+}\Delta$ respectively and consume
$\alpha_2{+}\Delta$ channel uses. However, the second virtual
channel is changed by reallocating the power of $u_4$ from
$P^{1{-}\Delta}$ to $P^{\alpha_2}$. In this way, the amount of
channel uses in this virtual channel is kept to
$2{-}\Delta{-}\alpha_2$, while the rates achieved become
$R_{u_4}{=}\alpha_2$ and $R_{\mathbf{v}_4}{=}2{-}\Delta{-}\alpha_2$.
As a result, the \emph{DoF} achieved using this 2 channel uses are
$\left(d_1{,}d_2\right){=}\left(\alpha_2{,}1\right)$.

\subsection{Achieving $\left(1,\alpha_1\right)$}

Point $\left(1,\alpha_1\right)$ is achieved in a ''SC+ZF'' manner
which has been mentioned in \cite{Ges12}. The transmission scheme is
finished in one slot and is expressed as
\begin{equation}
\mathbf{s}_t=\left[x_{c,t},0\right]^T+\hat{\mathbf{g}}_t^\bot
u_t+\hat{\mathbf{h}}_t^\bot v_t,\label{eq:sczf}
\end{equation}
where $x_{c,t}$ is intended to user 1 and transmitted with power $P$
superposed to $u_t$ and $v_t$, which are transmitted with power
$P^{\alpha_1}$. $u_t$ and $v_t$ are precoded to the orthogonal space
of the channel of their unintended user. At each receiver, $x_{c,t}$
is decoded with the rate of $1{-}\alpha_1$ using SIC. After
subtracting $x_{c,t}$, user 1 decodes $u_t$ with the rate of
$\alpha_1$ without any interference. Similarly, user 2 obtains $v_t$
with \emph{DoF} of $\alpha_1$. Resultantly, point
$\left(1,\alpha_1\right)$ is achieved.

As a last remark, we note the similarity of this work with
\cite{Chen12}, where the same \emph{DoF} region is derived under the
same system settings. Their achievable scheme in terminated in three
phases. The overheard interferences generated in current phase are
accumulated and then split evenly across the transmission in the
next phase. The optimal bound is achieved by choosing phase
durations which optimally combine the transmission of overheard
interference and new symbols. However, in this contribution, we
achieved the optimal bound by keeping retransmitting overheard
interference to boost up the \emph{DoF} and making every channel use
employed efficiently.

\section{Conclusion}\label{conclusions}

The optimal \emph{DoF} region is derived for a
two-user MISO broadcast channel when the transmitter has perfect
delayed CSI and asymmetric partial current CSI. A novel transmission
scheme is motivated and shown to achieve that \emph{DoF} region.
The results boil down to \cite{Tse10,Ges12,Gou12} in the symmetric scenario.

\section*{Appendix-Brief Proof of Outer-Bound}

The converse proposed in \cite{Ges12} can be reused to prove the
optimal region given in Theorem \ref{DoF_theorem}. In that work,
genie-aided model is employed to construct a physically-degraded
channel. Skipping the same fundamental steps, the only difference
lies in deriving the following lower bound
\begin{align}
& \mathcal{E}_{\tilde{\mathbf{\phi}}}\left(\log\left(1+\sum_{i=1}^m\lambda_i|\hat{\phi}_i+\tilde{\phi}_i|^2\right)\right)\nonumber\\
\geq &\mathcal{E}_{\tilde{\mathbf{\phi}}_1}\left(\log\left(\lambda_1|\tilde{\phi}_1|^2\right)\right) \nonumber\\
\approx & \log \lambda_1\sigma_2^2=\left(1-\alpha_2\right)\log P.
\label{eq:lb}
\end{align}$\lambda_1{>}\lambda_2{>}{\cdots}{>}\lambda_m$ are eigenvalues of
$\hat{\mathbf{g}}$. $\hat{\phi}_i{+}\tilde{\phi}_i$ are eigenvalues
of $\mathbf{g}$, $\tilde{\mathbf{\phi}}$ is Gaussian distributed as
$\mathcal{N}\left(0{,}\sigma_2^2\right)$. Replacing \eqref{eq:lb}
into its corresponding equations in \cite{Ges12}, we have
\eqref{eq:r1r2}. When switching the role of the two users,
\eqref{eq:r2r1} is obtained.
\begin{align}
R_1+2R_2 & \leq\left(2+\alpha_2\right)\log P, \label{eq:r1r2}\\
2R_1+R_2 & \leq\left(2+\alpha_1\right)\log P. \label{eq:r2r1}
\end{align}
\bibliographystyle{IEEEtran}
\bibliography{icc13timebib}

\end{document}